\documentclass[12pt,a4paper]{article}
\usepackage[utf8]{inputenc}
\usepackage[T1]{fontenc}
\usepackage{cite}
\usepackage{hyperref}
\usepackage{dsfont}
\usepackage{graphicx}
\usepackage{color}
\usepackage{stmaryrd}
\usepackage[multiple]{footmisc}
\usepackage{here}
\usepackage{bbm}
\usepackage{amsmath}
\usepackage{amssymb}
\usepackage{lineno}
\usepackage{geometry}
\usepackage{dsfont}
\usepackage{mathtools}
\numberwithin{equation}{section}
\usepackage{authblk}
\usepackage{hyperref}
\usepackage[british]{babel}
\usepackage{avant}
\usepackage{tgtermes} 
\usepackage{mathptmx}

\advance\textheight\topskip
\hypersetup{pdfborder={0 0 0},colorlinks=true,urlcolor=blue,citecolor=blue,linkcolor=blue,linktocpage=true}

\date{}

\begin{document}

\title{\vskip-1truecm{\textsc{\Large{Gravitating Meron-like topological solitons in massive Yang--Mills theory and the Einstein--Skyrme model}}}\vskip.5truecm}


\author[]{\textbf{Marcelo Ipinza}
}

\affil[] {{\normalsize\it Instituto de F\'isica Pontificia Universidad Cat\'olica de Valpara\'iso}\\ { \normalsize\it Casilla 4059, Valpara\'iso, Chile}}

\author[]{\textbf{Patricio Salgado-Rebolledo}
}

\affil[] {{\normalsize\it School of Physics and Astronomy, University of Leeds}\\ { \normalsize\it Leeds, LS2 9JT, United Kingdom}}

\maketitle

\vskip1.5truecm

\begin{center}
\begin{minipage}{.9\textwidth}
\textsc{\textbf{Abstract}}. We show that Merons in $D$-dimensional Einstein--Massive--Yang--Mills theory can be mapped to solutions of the Einstein--Skyrme model. The identification of the solutions relies on the fact that, when considering the Meron ansatz for the gauge connection $A=\lambda U^{-1}dU$, the massive Yang--Mills equations reduce to the Skyrme equations for the corresponding group element $U$. In the same way, the energy-momentum tensors of both theories can be identified and therefore lead to the same Einstein equations. Subsequently, we focus on the $SU(2)$ case and show that introducing a mass for the Yang--Mills field restricts Merons to live on geometries given by the direct product of $S^3$ (or $S^2$) and Lorentzian manifolds with constant Ricci scalar. We construct explicit examples for $D=4$ and $D=5$. Finally, we comment on possible generalisations.

\end{minipage}
\end{center}

\newpage
\tableofcontents
\newpage

\section{ Introduction}

The study of massive vector fields dates back to the seminal paper of Proca \cite{Proca:1900nv}, in which the analogue of the Klein-Gordon equation for a spin-1 field was constructed.  The Proca field can be consistently coupled to gravity, leading to the Einstein-Proca theory \cite{Tauber:1969ba} and generalisations \cite{Heisenberg:2014rta,Allys:2015sht,Chowdhury:2018nfv,GallegoCadavid:2019zke}, which have received renewed interest in the last years due to applicability in cosmology and astrophysics (see for example \cite{Golovnev:2008cf,Jimenez:2009ai,Brito:2015pxa,Chagoya:2016aar,deFelice:2017paw} and references therein).

Massive Yang--Mills theory (MYM) is the non-Abelian generalisation of Proca theory \cite{Kunimasa:1967zza,Veltman:1968ki,Reiff:1969pq,Shizuya:1975ek,Allys:2016kbq}. Non-Abelian massive gauge fields can be found in many different contexts in physics, such as the description of the W and Z bosons in the Standard Model \cite{Donoghue:1992dd}, non-perturbative QCD \cite{slavnov1979application} and cosmological models \cite{Bento:1992wy,Dimastrogiovanni:2010sm,Gomez:2020sfz}, among others. Even though MYM theory is non-renormalisable \cite{Komar:1960af,Umezawa:1961igm,Boulware:1970zc}, the problem can be circumvented by either introducing a Higgs boson \cite{Cornwall:1974km}, by modifying the ghost sector of the theory \cite{Fradkin:1970ib,Curci:1976bt,Fukuda:1981yf} (this, however, can bring problems with unitarity \cite{Delbourgo:1987np}), or by considering Abelian gauge conditions \cite{Ellwanger:2002sj}. Another possibility to avoid the renormalisability issue is to consider MYM theory as an effective description of the massless theory, where the mass is dynamically generated after gauge fixing \cite{Serreau:2012cg,Verschelde:2001ia}. In the case of QCD, there is strong evidence that gluons develop an effective mass at the non-perturbative level, which is related to colour confinement \cite{Cornwall:1981zr,Stingl:1985hx,Philipsen:2001ip,Dudal:2003by,Sobreiro:2004us,Dudal:2004rx,Sorella:2006ax,Aguilar:2008xm,Oliveira:2010xc,Canfora:2013kma}. Furthermore, at finite temperature gluons acquire dynamical mass when considering thermal loop corrections. Colour screening produces an electric Debye mass in the gluon propagator pretty much in the same way as it happens for the photon propagator in QED \cite{Shuryak:1977ut}. Moreover, the existence of magnetic monopole solutions in QCD leads to a purely non-perturbative magnetic screening mass \cite{Gross:1980br}.

On the other hand, an effective description of the low-energy regime of QCD is given by the Skyrme model \cite{Skyrme:1962vh}, namely, a non-linear field theory that describes the interaction of pions. In this context, baryons emerge as classical configurations given by topological solitons called Skyrmions. As shown by Atiyah and Manton \cite{Atiyah:1989dq} the holonomies of Instantons in Yang--Mills theory lead, in good approximation, to static solutions of the Skyrme model. This has been later generalised to define Skyrmions starting from Instantons living in higher dimensional spaces, which has applications in bulk/boundary duality scenarios \cite{Son:2003et,Sakai:2004cn,Eto:2005cc}. Moreover, this construction can be applied to the BPS Skyrme model \cite{Sutcliffe:2010et}, where the equivalence between Skyrmions and Instanton holonomies becomes exact.
Since a dynamical mass naturally appears when studying non-perturbative effects of gauge theories, it is interesting to investigate the possibility of a similar relation between topological solitons in MYM theory and Skyrmions.

In this paper, we focus on a very particular kind of topological solitons, namely, Meron-like solutions in Einstein-MYM theory. Merons are classical singular solutions of Yang--Mills equations with half-integer topological charge, which play an important role in the color confinement problem \cite{Actor:1979in,Lenz:2003jp}\footnote{Also, for a recent analysis of Meron configurations in the context of holographic QCD, see \cite{Suganuma:2020jng}.}. Although they were originally studied in Euclidean Yang-Mills theory as \emph{half}-Instantons\footnote{Merons have also been introduced as \emph{half}-Skyrmions in quantum Halls systems \cite{rho2017multifaceted}.}, they have been considered in the Lorentzian case as well. When coupled to gravity, the singularity of Merons can be avoided, rendering them well-defined solutions. In the four-dimensional case, a Lorentzian $SU(2)$ Meron gauge field ansatz has been used to construct non-Abelian black holes in Einstein--Yang--Mills theory \cite{Canfora:2012ap}. As shown in \cite{Bueno:2014mea}, the Meron field in this case resembles the Wu-Yang monopole solution, which was also used in \cite{Huebscher:2007hj} to define \emph{black hedgehogs} in $N=2$ supersymmetric Yang--Mills theory. In five dimensions, and analogue $SU(2)$ Meron black hole has been constructed by means of the generalised hedgehog ansatz \cite{Canfora:2018ppu}\footnote{Similar gauge field anstaz have been studied in \cite{Bostani:2009zf} as generalised monopole solutions with $SO(2,1)$ gauge group.}. Moreover, Euclidean Meron-like solutions in Einstein--Yang--Mills theory have been studied in \cite{Hosoya:1989zn,Rey:1989th,Gupta:1989bs,Maldacena:2004rf,Canfora:2017yio,Betzios:2017krj}.

In this article, we construct Meron-like solutions in the Einstein--MYM theory. We will show that these solutions can be mapped to solutions of the Einstein-Skyrme model, whose properties have been exhaustively studied in the last years \cite{Luckock:1986tr,Droz:1991cx,Heusler:1992av,Bizon:1992gb,Shiiki:2005aq,Ayon-Beato:2015eca,Canfora:2017ivv,Tallarita:2017bks,Astorino:2018dtr,Canfora:2018isz,Canfora:2019sxn,Ayon-Beato:2019tvu}. In particular, we show that a Meron solution of the MYM equations leads to a gauge group element that solves the Skyrme equations provided the coupling constants of both theories are identified in a suitable way. We show that the same identification allows one to obtain the Skyrme energy-momentum tensor from the corresponding one in MYM theory. This shows that both systems admit the same solutions of the Einstein equations for the space-time metric. In the $SU(2)$ case, we construct explicit examples based on the generalised hedgehog ansatz, which define massive Merons as well as Skyrmions on manifolds given by the direct product of $S^3$ and geometries with constant Ricci scalar. We show that a similar construction can be made for $S^2$ by performing an identification on the three-sphere that allows to map the generalised hedgehog ansatz on $S^3$ into the standard hedgehog on the two-sphere.

The paper is organised as follows. In Section \ref{SkyrmeandMYM}, we briefly review the Einstein--MYM theory and the Einstein--Skyrme model and show how Meron-like gauge fields in the former theory can be mapped to solutions of the latter one. Section \ref{SU2} deals with the $SU(2)$ case and the construction of Meron solutions on product spaces of constant Ricci scalar and $S^3$ (or $S^2$). The Euclidean continuation of the solutions is also addressed. Finally, in Section \ref{conclusions}, we summarise the results, give some concluding remarks and elaborate on possible future directions.

\section{Skyrmions from massive Merons}
\label{SkyrmeandMYM}

In this section, we show how Merons in Einstein-MYM theory can be used to construct Skyrmions in the Einstein--Skyrme model. In order to do so, we start by briefly reviewing the main aspects of both theories and subsequently introduce the Meron gauge field anstaz, which leads to the identification of massive Merons and Skyrmions.

\subsection{Einstein--MYM theory}

The starting point is the Einstein--MYM system in $D$ dimensions. The gravitational action is given by the Einstein-Hilbert term
\begin{equation}\label{EHaction}
I_{\rm EH}=\frac{1}{16\pi G}\int d^D x ~\sqrt{-g}\left( R-2\Lambda\right)\,,
\end{equation}
where $R$ is the Ricci scalar, $G$ is the Newton constant and $\Lambda$ is the cosmological constant. We have also set $c=\hbar=1$. The action for the massive Yang-Mills field is defined as the non-abelian generalisation of Proca theory \cite{Kunimasa:1967zza,Veltman:1968ki,Reiff:1969pq,Shizuya:1975ek,Allys:2016kbq}, i.e.
\begin{equation}\label{MYM}
I_{\rm MYM}=\frac{1}{16\pi \gamma^{2}}\,\int d^{D}x~\sqrt{-g}\,
{\rm Tr}\left[  F^{\mu\nu}F_{\mu\nu}+2m^{2}A^{\mu}A_{\mu}\right] \,.
\end{equation}
Here $A_\mu$ is a connection taking values on a matrix Lie algebra with associated gauge group $\mathfrak G$, $F_{\mu\nu}=\partial_{\mu}A_{\nu}-\partial_{\nu}A_{\mu
}+\left[  A_{\mu},A_{\nu}\right]$ is the corresponding field strength, and $\gamma$ is the Yang--Mills coupling constant. The action of the Einstein--MYM theory then reads
\begin{equation}\label{EMYM}
I_{\rm Einstein-MYM}=I_{\rm EH}+I_{\rm MYM}\,.
\end{equation}
Denoting the infinitesimal generators of $\mathfrak G$ by $t_i$, which satisfy the Lie algebra commutation relations
\begin{equation}\label{Liealg}
\left[ t_i, t_j\right]= f_{\;\;ij}^{k}\, t_k \,,
\end{equation}
the gauge field can be written as
\begin{equation}
A_{\mu}= A_{\mu}^{i}\,t_{i} \,,
\end{equation}
while the corresponding curvature can be written in components as
\begin{equation}\label{curv}
F_{\mu\nu}=F_{\mu\nu}^i \, t_i \,,\qquad F_{\mu\nu}^i=\partial_\mu A_\nu^i-\partial_\nu A_\mu^i+f_{\;\;jk}^{i}A_\mu^j A_\nu^k \,.
\end{equation}

Varying \eqref{EMYM} with respect to $A_\mu$ leads to the MYM equations
\begin{equation}\label{MYMeq}
\nabla_{\nu}F^{\nu\mu}+\left[  A_{\nu},F^{\nu\mu}\right]  -m^{2}%
A^{\mu}=0\,,
\end{equation}
where $\nabla^{\mu}$ is the Levi-Civita covariant derivative. One can see that evaluating the divergence of this equation leads to the condition
\begin{equation}\label{nablaA}
    \nabla_\mu A^\mu =0 \,.
\end{equation}
Thus, one of the consequences of introducing a mass term for the Yang-Mills field is that is that the solutions of \eqref{MYMeq} are divergence-free.

On the other hand, varying \eqref{EMYM} with respect to the metric leads to the $D$-dimensional Einstein equations
\begin{equation}\label{EeqMYM}
G_{\mu\nu}+\Lambda g_{\mu\nu}=8\pi G\, T^{\rm MYM}_{\mu\nu} \,,
\end{equation}
where $G_{\mu\nu}=R_{\mu\nu}-\frac{1}{2}g_{\mu\nu}R$ is the Einstein tensor and $T_{\mu\nu}$ is the stress-energy tensor associated to the MYM field:
\begin{equation}\label{tmunuMYM}
T^{\rm MYM}_{\mu\nu}
=\frac{1}{4\pi \gamma^{2}}\,{\rm Tr}\left[  -F_{\mu\alpha}F_{\nu\beta}%
g^{\alpha\beta}+\frac{1}{4}g_{\mu\nu}F^{\alpha\beta}F_{\alpha\beta}%
-m^{2}\left(A_{\mu}A_{\nu}-\frac{1}{2}g_{\mu\nu}A^{\alpha}A_{\alpha}\right)\right] \,.
\end{equation}
Note that, due to the presence of the mass term $m^2A^\mu A_\mu$, the Einstein-MYM action is not gauge invariant\footnote{Nevertheless, as it happens in the case of a Proca field, gauge symmetry can be restored in the theory by means of the St\"uckelberg mechanism \cite{Kunimasa:1967zza,Shizuya:1975ek}}.

\subsection{The Einstein--Skyrme model}
The Einstein-Skyrme model in $D$ dimensions is described by the action \cite{Luckock:1986tr,Droz:1991cx,Heusler:1992av,Bizon:1992gb,Shiiki:2005aq,Ayon-Beato:2015eca,Canfora:2017ivv,Tallarita:2017bks,Astorino:2018dtr,Canfora:2018isz,Canfora:2019sxn,Ayon-Beato:2019tvu}:
\begin{equation}\label{actionESm}
I_{\rm Einstein-Skyrme}= I_{\rm EH}+I_{\rm Skyrme}\,.
\end{equation}
The Einstein-Hilbert term $I_{\rm EH}$ has been defined in \eqref{EHaction}, whereas $I_{\rm Skyrme}$ is the Skyrme action for a matrix scalar field $U$, defined as a smooth map from space-time to a compact semi-simple matrix Lie group $\mathfrak G$. The Skyrme action reads  \cite{Skyrme:1962vh}
\begin{equation}\label{Skyrmeaction}
I_{\rm Skyrme}=\int d^D x \sqrt{-g}\,{\rm Tr}\left[\frac{f_{\pi}^{2}}{16}R_{\mu}R^{\mu}+\frac{1}{32e^{2}}H_{\mu\nu}H^{\mu\nu}\right] \,,
\end{equation}
where $f_\pi$ and $e$ are coupling constants to be determined experimentally \cite{Adkins:1983ya}, and we have defined
\begin{equation}\label{defRandH}
R_\mu=U^{-1}\partial_\mu U \,,\hskip 1.5truecm H_{\mu\nu}=\left[R_\mu , R_\nu \right]  \,.  
\end{equation}
Varying the action \eqref{actionESm} with respect to $U$ leads to Skyrme equation
\begin{equation}\label{skyrme}
\nabla_{\mu}R^{\mu}+\frac{1}{e^{2}f_{\pi}^{2}}\nabla_{\mu}\left[R_{\nu},H^{\mu\nu}\right]=0 \,,
\end{equation}
whereas variation with respect to the metric yields the Einstein equations
\begin{equation}\label{EeqSkyrme}
    G_{\mu\nu}+\Lambda g_{\mu\nu}=8\pi G \, T^{\rm Skyrme}_{\mu\nu} \,,
\end{equation}
where the energy-momentum tensor of the Skyrme model is given by
\begin{equation}\label{tmunuS}
T_{\mu\nu}^{\rm Skyrme}=
-\frac{1}{8}{\rm Tr}\left[f_{\pi}^{2} \left(R_{\mu}R_{\nu}-\frac{1}{2}g_{\mu\nu}R^{\alpha}R_{\alpha}\right)+\frac{1}{e^{2}}\left(H_{\mu\alpha}H_{\nu}^{\;\;\alpha}-\frac{1}{4}g_{\mu\nu}H_{\alpha\beta}H^{\alpha\beta}\right)\right] \,.
\end{equation}
Skyrmions are characterised by their topological charge, given by the Baryon number
\begin{equation}\label{bnumber}
    B=\frac{1}{24\pi^2} \int {\rm Tr}\left[R_\mu R_\nu R_\rho\right] dx^\mu \wedge dx^\nu \wedge dx^\rho  \,,
\end{equation}
where integration is to be made over a three-dimensional compact hypersurface on which $R_\mu$ has support at spatial infinity.

\subsection{Translating massive Merons into Skyrmions}\label{MYMtoS}
In this section, we will show how Meron gauge fields constructed in the Einstein--MYM theory can be translated into solutions of the Einstein--Skyrme model. 

In Yang--Mills theories, Meron configurations are defined as gauge potentials that are proportional to a pure gauge \cite{Actor:1979in}, i.e.
\begin{equation}\label{Ameron}
A_{\mu}=\lambda U^{-1}\partial_{\mu}U \,,\hskip 1.5truecm\lambda\neq0,1 \,.
\end{equation}
In the case of MYM theory, pure gauge configurations do not exist as the presence of the mass $m$ in \eqref{MYM} breaks the gauge symmetry of the Yang--Mills ($m=0$) case. However, as it will be shown in Section \ref{SU2}, it is possible to find solutions of the MYM equations for gauge fields of the form \eqref{Ameron} for a restricted class of product space manifolds. In the limit $m\rightarrow0$ these solutions reduce to Meron solutions of the standard (massless) Einstein--Yang--Mills theory\footnote{Some of the solutions in the $m=0$ case have been already constructed in \cite{Canfora:2017yio} in the Euclidean case.}.

Note that the definition \eqref{Ameron} leads to a curvature form that is algebraic in the gauge potential, i.e.
\begin{equation}\label{Fmeron}
F_{\mu\nu}=\frac{\lambda-1}{\lambda} \left[A_\mu,A_\nu\right] \,.
\end{equation}
Therefore, given an element $U\in\mathfrak G$ that defines a Meron gauge field \eqref{Ameron}, one can write
\begin{equation}\label{meronAF}
A_\mu=\lambda R_\mu \,,\qquad F_{\mu\nu}=\lambda\left(\lambda-1\right)H_{\mu\nu} \,,
\end{equation}
where we have intentionally adopted the notation \eqref{defRandH} defined for the Einstein--Skyrme model. 

Consider now the Einstein equations in the MYM case \eqref{EeqMYM}. The energy-momentum tensor entering in this equation is given by \eqref{tmunuMYM}, which in the case of a Meron gauge field \eqref{meronAF} reduces to
\begin{equation}
\begin{aligned}
    T^{\rm MYM}_{\mu\nu}\Big|_{A_\mu=\lambda R_\mu}&=-{\rm Tr}\Bigg[\frac{\lambda^2\left(1-\lambda\right)^2}{4\pi \gamma^2} 
    \left(H_{\mu\alpha}H_{\nu}^{\;\;\alpha}-\frac{1}{4}g_{\mu\nu}H_{\alpha\beta}H^{\alpha\beta}\right)\\
  & \hskip 4.8 truecm +\frac{m^2\lambda^2}{4\pi \gamma^2}
\left(R_{\mu}R_{\nu}-\frac{1}{2}g_{\mu\nu}R^{\alpha}R_{\alpha}\right)
    \Bigg]\,.
\end{aligned}
\end{equation}
It is straightforward to see that this expression corresponds to the energy-momentum tensor of the Skyme model \eqref{tmunuS}, provided we identity the constants $\lambda$, $m$ and $\gamma$ in the form
\begin{equation}\label{idparameters}
    \frac{m^2\lambda^2}{\pi \gamma ^2}=\frac{f_\pi^2}{2}\,,\qquad \frac{\lambda^2 \left(\lambda-1\right)^2}{\pi \gamma ^2}=\frac{1}{2e^2}\,. 
\end{equation}
This means that, finding a Meron gauge field $A_\mu=\lambda R_\mu $ that solves Einstein  equations \eqref{EeqMYM} in the Einstein--MYM system, allows one to find a group element $U\in\mathfrak G$ that solves the corresponding Einstein equations \eqref{EeqSkyrme} in the Einstein--Skyrme model. It is important to remark that the relation \eqref{idparameters} rules out the case $m=0$ and therefore it cannot be established in the case of (massless) Yang--Mills theory. Furthermore, it requires the condition $\lambda\neq 0,1$ to hold, as specified in \eqref{Ameron}, which ensures that the field strength tensor \eqref{Fmeron} does not vanish.  

To complete this identification, we still have to show that a Meron solution of the MYM equations determines a solution of the Skyrme equations. In order to do so, let us replace the ansatz \eqref{meronAF} in the MYM equations \eqref{MYMeq}. This leads to
\begin{equation}\label{RHYMeq}
\nabla_\nu H^{\nu\mu} +\lambda\left[R_\nu,H^{\nu\mu}\right]-\frac{m^2}{\lambda-1}R^\mu=0\,.
\end{equation}
Either evaluating the commutator of this equation with $R_\mu$ or acting on it with $\nabla_\mu$, leads to
\begin{equation}\label{commRskyrme}
\nabla_\mu\left[R_\nu,H^{\mu\nu}\right]=0 \,.
\end{equation}
On the other hand, the condition \eqref{nablaA} in this case simply means
\begin{equation}\label{nablaR}
\nabla_\mu R^\mu =0\,.
\end{equation}
From \eqref{nablaR} and \eqref{commRskyrme} we see that a Meron satisfying the MYM equations  \eqref{MYMeq} defines a group element $U\in \mathfrak G$ satisfying the Skyrme equation \eqref{skyrme}. This completes the proof that gravitating Merons in Einstein--MYM theory can be mapped to solutions of the Einstein--Skyrme model.

It is interesting to note that the Skyrme action can also be recovered from the MYM one. Plugging the ansatz \eqref{meronAF} back into the MYM action \eqref{MYM} yields the \emph{effective} reduced action 
\begin{equation}
I_{\rm MYM}\Big|_{A_\mu=\lambda R_\mu}=\frac{1}{16\pi \gamma^2}\int d^D x \, \sqrt{-g} \,{\rm Tr}\,\left[\lambda^2\left(1-\lambda\right)^2H_{\mu\nu}H^{\mu\nu}+2m^2\lambda^2 R_{\mu}R^\mu\right] \,,
\end{equation}
which corresponds to the Skyrme action \eqref{Skyrmeaction} provided we implement \eqref{idparameters}.

\section{$SU(2)$ case and product manifolds}
\label{SU2}
We will now consider the particular case $\mathfrak G = SU(2)$ and define Meron gauge field configurations following \cite{Canfora:2012ap,Canfora:2017yio}. We will consider the generators of the $\mathfrak{su}(2)$ algebra in the form
\begin{equation}
t_{i}=i\sigma_{i}\,,\hskip 0.5truecm i=1,2,3\,,
\end{equation}
where $\sigma_{i}$ stands for the Pauli matrices. Therefore, Eqs. \eqref{Liealg} and \eqref{curv} hold for
\begin{equation}
    f_{ijk}=-2\,\epsilon_{ijk} \,,
\end{equation}
where $\epsilon_{ijk}$ is three-dimensional the Levi--Civita symbol. Note that a pure gauge field can be expressed in terms of SU(2) left-invariant Maurer--Cartan forms 
\begin{equation}\label{formR}
R=R_\mu dx^\mu =U^{-1}dU=i\Gamma^i \sigma_i \,,
\end{equation}
which satisfy the Maurer--Cartan equation
\begin{equation}\label{MCeq}
    d\Gamma^i=\epsilon^i_{\;\;jk}\Gamma^j\wedge \Gamma^k \,,
\end{equation}
Therefore, one can define the Meron one-form using \eqref{Ameron} and \eqref{formR},
\begin{equation}\label{Aform}
    A=A_\mu dx^\mu =i\lambda \Gamma^i \sigma_i \,,
\end{equation}
and similarly for the corresponding curvature two-form, for which Eq. \eqref{Fmeron} leads to
\begin{equation}\label{Fform}
    F=\frac{1}{2}F_{\mu\nu}dx^\mu dx^\nu = -i\lambda\left(\lambda-1\right)\epsilon^{i}_{\;\;jk}\,
    \Gamma^j \wedge \Gamma ^k 
    \sigma_i \,.
\end{equation}
In the following, we will use these expressions to evaluate the MYM equations \eqref{MYMeq} and solve them to find a solution for $\lambda$.

\subsection{Metric ansatz and massive Yang--Mills equation}\label{metricansatz}
Let us consider a $D$-dimensional space-time with line element of the form \cite{Canfora:2017yio}
\begin{equation}\label{metric}
ds^{2}=g_{\mu\nu}dx^{\mu}dx^{\nu}=h_{ab}\left(  z\right)  dz^{a}%
dz^{b}+\rho(z)^{2}d\Omega^2_3\left(y\right)  \,,
\end{equation} 
where we have split the coordinates as
\begin{equation}
x^{\mu}=\left(z^{a},y^{i}\right) \,,
\hskip .5truecm
a=0,..,D-4\,, 
\hskip .5truecm
i=1,2,3\,.
\end{equation}
Here, $h_{ab}$ is a $(D-3)-$dimensional metric expressed in coordinates $z^a$, $d\Omega^2_3$ is the line element of the three-sphere $S^3$ with coordinates $y^i$, and $\rho\left(z \right)$ is a warping factor depending only on the coordinates $z^{a}$. We will use the bi-invariant metric on $SU(2)$ \cite{Milnor:1976pu} to define the metric for $S^3$ in terms of the Maurer-Cartan forms $\Gamma^i$:
\begin{equation}
d\Omega_3=-\frac{1}{2}{\rm Tr}\left[U^{-1}dU \, U^{-1}dU\right]=\sum_{i=1}^{3}\Gamma^{i}\,\Gamma^{i} \,.
\end{equation}
Therefore, one can define a set of vielbeins $e^A=\left(e^{\tilde A},e^{D-4+i}\right)$ for \eqref{metric}, where $\tilde A=0,\dots,D-4$, such that
\begin{equation}
ds^2=\eta_{AB}e^A \, e^B \,,
\hskip .5truecm 
h_{ab}\left(z \right) dz^a dz^b=\eta_{\tilde A \tilde B}\,e^{\tilde A} e^{\tilde B} \,,
\hskip .5truecm 
\rho^2 d\Omega^2_3\left(y\right)= \sum_{i=1}^{3} e^{D-4+i}\, e^{D-4+i} \,.
\end{equation}
In particular, this means that the Maurer--Cartan forms \eqref{MCeq} define a set of dreibeins on $S^3$ and thus
\begin{equation}\label{dreibeinsS3}
    e^{D-4+i}=\rho\left(z\right)\,\Gamma^i \left(y\right)\,,
\end{equation}
where the forms $\Gamma^i$ are to be expressed in terms of the coordinates $y^i$ on $S^3$. The relation \eqref{dreibeinsS3} makes explicit the identification between tangent space-time indices and Lie algebra indices $i,j,k$. This identification has been used in \cite{Canfora:2018ppu} to define the spin-from-isospin effect for gravitating Merons.

As we will see now, the previous considerations are sufficient to solve the MYM equations \eqref{MYMeq}, which can be written in differential form language as
\begin{equation}\label{MYMeqforms}
    d*F+[A,*F]+m^{2}*A=0\,,
\end{equation}
Using \eqref{dreibeinsS3}, one can show that the Hodge dual of the one-form Meron gauge field \eqref{Aform} and two-form curvature \eqref{Fform} read
\begin{equation}\label{hodges}
\begin{aligned}
*A &=(-1)^{D-1}\,\frac{i\lambda\rho\left(z\right)}{2}d\Gamma^{i}\,\sigma_{i} \wedge{\rm Vol}_h\,,
\\
*F &= (-1)^{D}\,\frac{2i\lambda\left(\lambda-1\right)}{\rho\left(z\right)}\Gamma^{i}\,\sigma_{i}\wedge {\rm Vol}_h \,,
\end{aligned}
\end{equation}
where ${\rm Vol}_h = e^{0}\wedge\cdots\wedge e^{D-4}$ is the volume form associated to the $(D-3)$-dimensional manifold defined by the metric tensor $h_{ab}(z)$ in \eqref{metric}. Using these expression and the Maurer--Cartan equation \eqref{MCeq}, all the components of the MYM equation \eqref{MYMeqforms} can be shown to be proportional to a single polynomial equation for $\lambda$:
\begin{equation}
    (-1)^{D}\frac{i\lambda}{\rho\left(z\right)}\left(4\lambda\left(\lambda-1\right)-2\left(\lambda-1\right)+\frac{m^{2}\rho\left(z\right)^{2}}{2}\right)d\Gamma^{i}\,\sigma_{i}\wedge {\rm Vol}_h =0 \,,
\end{equation}
which can be solved for
\begin{equation}\label{rhoconstant}
    \rho\left(z\right)=\rho_0={\rm const.\neq0}
\end{equation}
and leads to
\begin{equation}\label{sollambda1}
    \lambda_\pm=\frac{1}{4}\left(3\pm \sqrt{1-2m^2 \rho_0 ^2}\right)\,.
\end{equation}
In the massless limit $m\rightarrow 0$, the solution $\lambda_- \rightarrow 1/2$ reproduces the usual result for Merons considered in \cite{Canfora:2017yio,Canfora:2018ppu}, while the solution $\lambda_+ \rightarrow 1$ becomes trivial. Note that \eqref{sollambda1} requires the mass to be bounded from above, i.e.
\begin{equation}\label{condm1}
    m^2\leq\frac{1}{2\rho_0^2} \,.
\end{equation}
The condition \eqref{rhoconstant} implies that, in order to have a Meron gauge field solving the MYM equations in a space-time background of the form \eqref{metric}, the space-time must be a product manifold of the three-sphere and a $(D-3)$-dimensional manifold $M_{D-3}$ with metric $h_{ab}(z)$.

As a consistency check, it is worth to note that we can use \eqref{hodges} to evaluate the exterior derivative of $*A$, which takes the form
\begin{equation}
    d*A =(-1)^{D-1}\,\frac{i\lambda}{2}\frac{\partial \rho\left(z\right)}{\partial z^a}dz^a\wedge d\Gamma^{i}\,\sigma_{i} \wedge{\rm Vol}_h \,.
\end{equation}
Therefore the condition \eqref{nablaA}, which in this case is given by $*d*A=0$, is fulfilled for constant $\rho$ and matches the condition \eqref{rhoconstant} found from solving the MYM equations. 

\subsection{Generalised hedgehog ansatz}
An element $U\in SU(2)$ can be parametrised in the form 
\begin{equation}
U\left(  x^{\mu}\right)  =\cos\alpha\left(  x^{\mu}\right)  \mathbbm{1}+   i \sin\alpha\left(x^{\mu}\right)  n^{i}\sigma_{i} \,,
\end{equation}
where $\mathbbm{1}$ is the $2\times2$ identity and $n^i$ is a unitary vector. In this parametrisation, the components of the left-invariant Maurer--Cartan forms on $SU(2)$ \eqref{formR} are given by
\begin{equation}\label{Gmammasgen}
    \Gamma^i=\sin\alpha\, \epsilon^{i}_{\;jk}n^j dn^k+ n^i d\alpha +\sin\alpha\cos\alpha\, dn^i \,,
\end{equation}
The generalised hedgehog ansatz \cite{Canfora:2013mrh,Canfora:2013osa} consists in choosing
\begin{equation}\label{unitvector}
n^{1} =\sin\Theta\cos\Phi\,,\hskip .5truecm
n^{2}=\sin\Theta\sin\Phi\,,\hskip .5truecm
n^{3}=\cos\Theta\,,
\end{equation}
where $\Phi$, $\Theta$ are functions of the space-time coordinates. In this case we will consider the ansatz introduced in \cite{Ayon-Beato:2015eca} (see also \cite{Canfora:2017ivv,Canfora:2017yio,Astorino:2018dtr,Canfora:2018ppu,Canfora:2019sxn,Ayon-Beato:2019tvu}):
\begin{equation} \label{gha}
\Phi=\frac{\psi+\phi}{2}\,, \hskip 1.2truecm 
\tan\Theta =\frac{\cot\left(  \frac{\theta}{2}\right)}{\cos\left(  \frac{\psi-\phi}{2}\right)} \,,\hskip 1.2truecm 
\tan  \alpha  =-\frac{\sqrt{1+\tan^{2}  \Theta }}{\tan\left(  \frac{\psi-\phi}{2}\right)  } \,, 
\end{equation}
where $0<\psi<4\pi$, $0<\theta<\pi$, $0<\phi<2\pi$. This leads to\footnote{One can see that the generalised hedgehog ansatz consist in expressing the $SU(2)$ left-invariant forms in terms of Euler angles \cite{varshalovich1988quantum}.}
\begin{equation}\label{GammasEuler}
\begin{aligned}
\Gamma_{1}  &  =\frac{1}{2}\left(  \sin\psi \, d\theta
-\sin\theta  \cos\psi \,d\phi\right) \,,\\
\Gamma_{2}  &  =\frac{1}{2}\left( -\cos\psi \,d\theta
-\sin\theta \sin\psi \,d\phi\right) \,, \\
\Gamma_{3}  &  =\frac{1}{2}\left(  d\psi+\cos\theta\,
d\phi\right) \,.
\end{aligned}
\end{equation}
Using this expression in \eqref{Aform}, one can verify that for any metric of the form in \eqref{metric} a Meron gauge field satisfies the condition \eqref{nablaA}. It is important to note that the generalised hedgehog ansatz is topologically non-trivial and thus cannot be deformed continuously to the trivial vacuum. Indeed, it has non-trivial winding number along the $z^{a}={\rm const}$ hyper-surfaces of the metric \eqref{metric},
\begin{equation}\label{winding}
W=\frac{1}{24\pi^{2}}\int_{S^{3}}{\rm Tr}\left[  \left(  U^{-1}dU\right)
^{3}\right]  =1\,.
\end{equation}
This expression corresponds to the Baryon charge for the Skyrme field $U$ given by \eqref{bnumber}. Thus, massive gauge fields constructed by means of the generalised hedgehog anstaz are characterised by $B=1$.

\subsection{Einstein equations}
In Section \ref{metricansatz}, we have shown that Meron solutions in MYM theory with gauge $SU(2)$ exist on products manifolds of the form $M_{D-3}\times S^3$ with metric
\begin{equation}\label{prodmanifold}
ds^{2}=h_{ab}\left(z\right)  dz^{a}%
dz^{b}+\frac{\rho_0^{2}}{4}\left(d\psi^{2}+d\theta^{2}+d\phi^{2}+2\cos\theta d\phi d\psi\right) \,,
\end{equation}
where we have used \eqref{metricS3Euler} to express the line element of the three-sphere $d\Omega^2_3$. For such kind of spaces, the components of the Einstein tensor are given by \cite{Canfora:2017yio}
\begin{equation}
\begin{aligned}
&G_{\psi\psi}  = G_{\theta\theta}=G_{\phi\phi}=\frac{1}{\cos\theta}G_{\psi\phi}=-\frac{1}{4}\left(1+\frac{\rho_0^{2}}{2}R^{(h)}\right)\,, \\
&G_{ia} =0\,,\\
&G_{ab} =  G^{(h)}_{ab}-\frac{3}{\rho_0^{2}}h_{ab}\,,
\end{aligned}
\end{equation}
where $R^{(h)}$ and $G^{(h)}_{ab}$ are the Ricci scalar and the Einstein tensor associated to the metric $h_{ab}$, respectively. The energy-momentum tensor \eqref{tmunuMYM}, on the other hand, can be evaluated using \eqref{Aform} and \eqref{GammasEuler}, which yields
\begin{equation}
\begin{aligned}
&T_{\psi\psi}  = T_{\theta\theta}=T_{\phi\phi}=\frac{1}{\cos\theta}T_{\psi\phi}=\frac{\lambda^2}{4\pi\gamma^2 }
\left(\frac{(\lambda -1)^2}{\rho^2_0}-\frac{m^2}{4} \right)\,, \\
&T_{ia} =0\,,\\
&T_{ab} = - \frac{3\lambda^2}{\pi\gamma^2 }\left(\frac{(\lambda -1)^2}{\rho^4_0}+\frac{m^2}{4\rho^2_0}\right) h_{ab}\,.
\end{aligned}
\end{equation}
Using these results, the Einstein equations \eqref{EeqMYM} reduce to
\begin{equation}\label{effEqs}
\begin{aligned}
 &R^{(h)}=R_0 \,,\\
 &G^{(h)}_{ab}+\Lambda_{\rm eff} \,h_{ab}=0 \,,
\end{aligned}
\end{equation}
where we have defined the constants
\begin{equation}\label{effEqsN2}
\begin{aligned}
R_0&=2\Lambda-\frac{2(\gamma^2-2Gm^2\lambda^2)}{\gamma^2\rho_0^2}-\frac{16G\lambda^2(\lambda -1)^2}{\gamma^2 \rho_0^4} \,,
 \\
\Lambda_{\rm eff}&=\Lambda-\frac{3(\gamma^2-2Gm^2\lambda^2)}{\gamma^2\rho_0^2}+\frac{24G\lambda^2(\lambda -1)^2}{\gamma^2 \rho_0^4}\,.
\end{aligned}
\end{equation}
As shown in Section \ref{MYMtoS}, the energy-momentum tensor in Einstein--MYM theory \eqref{tmunuMYM} and the energy-momentum tensor of the Einstein--Skyrme model \eqref{tmunuS} can be identified by replacing $m$ and $\gamma$ by the constants $f_\pi$ and $e$ according to \eqref{idparameters}. Therefore, when implementing \eqref{idparameters}, Eq. \eqref{effEqs} also defines the Einstein equations of the Einstein--Skyrme model \eqref{EeqSkyrme} evaluated on product manifolds of the form \eqref{prodmanifold}. In both cases, the energy the solutions is given by
\begin{equation}\label{energy}
\begin{aligned}
E&= \int_{t = \rm const.} d^{D-1}x \sqrt{-g} \;T_{00}\\
&=- \frac{6\pi\lambda^2}{\gamma^2 }\left(\frac{(\lambda -1)^2}{\rho_0}+\frac{m^2\rho_0}{4}\right) \int d^{D-4}z  \sqrt{-h} \;h_{00}(z)\,.
\end{aligned}
\end{equation}

In the following, we will show examples of solutions of these equations, which by construction define space-time backgrounds that support massive gravitating Merons and gravitating Skyrmions.

\subsubsection*{D>5}
For $D>5$, the solutions of the Einstein equations \eqref{effEqs} are product manifolds of $S^3$ and $(D-3)$-dimensional geometries $M_{D-3}$ with constant Ricci scalar $R_0$, which in turn satisfy a set of Einstein equations in $D-3$ dimensions with an effective cosmological constant $\Lambda_{\rm eff}$.
\subsubsection*{D=5}
The case $D=5$ is special since the submanifold defined by the metric $h_{ab}$ has dimension two and the associated Einstein tensor $G_{ab}^{(h)}$ vanishes identically. In this case, the manifold $M_2$ endowed with the $(1+1)$-dimensional metric $h_{ab}$ is a constant curvature space with Riemann tensor
\begin{equation}
    R^{(h)}_{abcd}=\frac{R_0}{2}
    \left(h_{ac}h_{bd}-h_{ad}h_{bc}\right) \,.
\end{equation}
The second equation in \eqref{effEqs} reduces to the condition $\Lambda_{\rm eff}=0$, which leads to the following solution for $\rho_0^2$ when $\Lambda\neq0$:
\begin{equation}\label{solrhopm}
    \rho^2_{0\pm}=\frac{3\left(\gamma^2-2Gm^2\lambda^2 \right) \pm\sqrt{9\left(\gamma^2-2Gm^2\lambda^2\right)^2-96\gamma^2G\Lambda \lambda^2\left(\lambda-1\right)^2}}{2\gamma^2\Lambda} \,.
\end{equation}
For the manifold $M_2$ with metric $h_{ab}$ we can consider coordinates $z^a=(t,r)$ and the general ansatz
\begin{equation}\label{metrich}
h_{ab}dz^a dz^b=-f(r)^2\,dt^2+\frac{1}{f(r)^2}\,dr^2 \,.
\end{equation}
The Ricci scalar associated to this metric has the form
\begin{equation}
R^{(h)}= -\left(f(r)^2\right)^{\prime\prime}\,,
\end{equation}
which together with Eq. \eqref{effEqs} leads to
\begin{equation}\label{fr}
   f(r)^2=-\frac{R_0}{2} r^2+C_2 r +C_1\,,
\end{equation}
where $C_1$ and $C_2$ are arbitrary constants. 
The case $C_1=1$, $C_2=0$ corresponds to the product manifold dS$_2\times S^3$ when $R_0>0$ and to AdS$_2\times S^3$ when $R_0<0$. Since $\rho_0^2$ must be positive, it is necessary to consider the different possible choices for the cosmological constant $\Lambda$ separately. For $\Lambda>0$ both $\rho^2_{0+}$ and $\rho^2_{0-}$ are admissible solutions and the following conditions must hold
\begin{equation}\label{5cond1}
     \gamma^2>2Gm^2 \lambda^2\,,\hskip 1.2truecm
     \Lambda<\frac{3\left(\gamma^2-2G m^2 \lambda^2\right)^2}{32G\gamma^2\lambda^2\left(\lambda-1\right)^2} \,.
\end{equation}
For $\Lambda<0$ the solution $\rho^2_{0+}$ is ruled out, while $\rho^2_{0-}$ is always well-defined. In the case $\Lambda=0$ the solution has to be worked out independently, and the vanishing of $\Lambda_{\rm eff}$ yields
\begin{equation}\label{rhozero}
    \rho_0^2=\frac{8G\lambda^2\left(\lambda-1\right)^2}{\gamma^2-2G m^2 \lambda^2}\text{ ,}
\end{equation}
which requires the first condition of \eqref{5cond1} to hold\footnote{Note that we have expressed $\rho_0^2$ in terms of $\lambda$, while the opposite has been done in \eqref{sollambda1}. When plugging \eqref{solrhopm} or \eqref{rhozero} in \eqref{sollambda1}, it is possible to express $\lambda$ as a quite involved function of $m$, $\gamma$, $G$ and $\Lambda$, whose explicit form will not be given here.}. As we can see from \eqref{energy}, in this case the energy is given by
\begin{equation}
 E= \frac{6\pi\lambda^2}{\gamma^2 }\left(\frac{(\lambda -1)^2}{\rho_0}+\frac{m^2\rho_0}{4}\right) \int_0^\infty dr f(r)^2 \,,
\end{equation}
which clearly diverges. We can easily see from this expression that in the massless case, if instead of a constant $\rho_0$ we had $\rho=r$ and also $f(r)^2=1$, we would have obtained the usual logarithmically divergent energy for Merons on flat space. This case is ruled our from our discussion since it does not satisfy \eqref{rhoconstant}. Moreover, from this result we see that going to four dimensions we can find a solution with finite energy.

\subsubsection*{D=4}
In four dimensions the only possible Lorentzian geometry that can be considered is $\mathbb R\times S^3$. In this case $z^a=t$ and we can choose $h_{ab}=h_{11}=1$. Naturally, in this case $R^{(h)}=0=G_{ab}^{(h)}$ and thus $R_0=0=\Lambda_{\rm eff}$. Solving \eqref{effEqsN2} fixes the cosmological constant to be
\begin{equation}\label{lambdaRxS}
\Lambda=\frac{1}{\rho_0^2}-\frac{2G\lambda^2 m^2 }{\gamma^2\rho_0^2}+\frac{8G\lambda^2 (\lambda-1)^2}{\gamma^2\rho_0^4}\,,
\end{equation}
whereas $\rho_0$ takes the form
\begin{equation}\label{rhoRxS}
\rho_0^2=\frac{16G\lambda^2(\lambda -1)^2}{\gamma^2-2Gm^2\lambda^2}\,.
\end{equation}
The energy of the solution is finite and has the form
\begin{equation}\label{finiteE}
 E= \frac{6\pi\lambda^2}{\gamma^2 }\left(\frac{(\lambda -1)^2}{\rho_0}+\frac{m^2\rho_0}{4}\right) \,,\end{equation}
and therefore this configuration corresponds to massive Meron or, equivalently, to a Skyrmion with finite energy, which was first found in \cite{Ayon-Beato:2015eca}. As we will see in Section \ref{Euc}, the corresponding Euclidean configuration has vanishing on-shell action.

In the following, we will show how similar solutions can be constructed on product spaces of $(D-2)$-dimensional manifolds and the two-sphere by performing a coordinate identification in the generalised hedgehog ansatz as well as in the metric of the three-sphere.

\subsection{From $M_{D-3} \times S^3$\, to\, $M_{D-2} \times S^2$}
Since the intersection of $S^3$ with a three-dimensional plane defines a two-sphere, a coordinate identification of the three- sphere allows one to reduced the generalised hedgehog anstaz \eqref{gha} to $S^2$. The metric of the three-sphere is given by
 \begin{equation}\label{metricS3Euler}
d\Omega^2_{3}=\frac{1}{4}\left(d\psi^{2}+d\theta^{2}+d\phi^{2}+2\cos\theta \,d\phi d\psi\right) \,,
 \end{equation}
One can see that, defining new variables $(\vartheta,\varphi)$ and setting
\begin{equation}
\psi=\phi=\varphi\,,
\hskip 1.5truecm
\theta=\pi-2\vartheta \,,
\end{equation} 
the line element \eqref{metricS3Euler} reduces to the line element of the two-sphere
\begin{equation}\label{idangles}
d\Omega^2_{3}\rightarrow d\Omega^2_{2}= d\vartheta^{2}+\sin^{2}\vartheta\, d\varphi^{2} \,.
\end{equation}
However, as $\theta\in[0,\pi)$, the new angle $\vartheta=(\pi-\theta)/2$ ranges over $[\frac{\pi}{2},0)$. Thus, the identification \eqref{idangles} has to be supplemented with the range redefinition\footnote{This in complete analogy with the fact that identifying the polar angle and the azimuth on a two-sphere leads to a half-circle, whose resulting angle range must doubled to obtain $S^1$. The resulting circle is equivalent to an intersection of the two-sphere and a two-dimensional plane.}.
\begin{equation}
0<\varphi<2\pi\,,
\hskip .5truecm
0<\vartheta<\pi\,.
\end{equation} 
This identification allows one to define Merons on $S^2$. Indeed, replacing \eqref{idangles} in the generalised hedgehog anstaz \eqref{gha} yields
\begin{equation} \label{gha2}
\Phi=\varphi\,, \hskip 1.2truecm 
\Theta =\vartheta \,,\hskip 1.2truecm 
\alpha  =\frac{\pi}{2} \,. 
\end{equation}
Implementing \eqref{gha2} reduces \eqref{unitvector} to the usual unitary vector on $S^2$, 
\begin{equation}
    n^1=\sin\vartheta \cos\varphi\,,\hskip .5truecm
    n^2=\sin\vartheta\sin\varphi\,, \hskip .5truecm
    n^3=\cos\vartheta\,,
\end{equation}
and therefore turns the generalised hedgehog anstaz into the standard spherically symmetric hedgehog. The choice \eqref{gha2} has been used in \cite{Canfora:2012ap} to define Merons on $S^2$. Here we have shown how it can be obtained by an identification of the three-sphere. Now, applying \eqref{idangles} on \eqref{prodmanifold}, we can define a Meron on a product space given by a $(D-2)$-dimensional manifolds $M_{D-2}$ times $S^2$, i.e.
\begin{equation}\label{prodmanifold2}
ds^{2}=h_{ab}\left(z\right)  dz^{a}%
dz^{b}+\rho_0^{2}\left(d\vartheta^{2}+\sin^2\vartheta d\varphi\right)\,.
\end{equation}
When applied to the Maurer--Cartan forms \eqref{GammasEuler}, this leads to
\begin{equation}\label{IdGammasEuler}
\begin{aligned}
\Gamma_{1}  &\hskip .2truecm  \rightarrow  \hskip .2truecm
-\sin\varphi \, d\vartheta
-\cos\vartheta \sin\vartheta \cos\varphi\, d\varphi \,,\\
\Gamma_{2}  &\hskip .2truecm  \rightarrow \hskip .2truecm
 \cos\varphi\, d\vartheta
- \cos\vartheta \sin\vartheta \sin\varphi\, d\varphi \,,\\
\Gamma_{3}  &\hskip .2truecm  \rightarrow \hskip .2truecm
\sin^2\vartheta\,
d\varphi\,,
\end{aligned}
\end{equation}
which determines the Meron gauge field one-form \eqref{Aform} and the curvature \eqref{Fform}. Replacing these expressions into the MYM equations \eqref{MYMeq} leads to the following polynomial for $\lambda$
\begin{equation}
4\lambda\left(\lambda-1\right)-2\left(\lambda-1\right)+m^{2}\rho_0^{2} =0 \,.
\end{equation}
Therefore, instead of \eqref{sollambda1}, in the case of Merons on $S^2$ the solution for $\lambda$ is given by
\begin{equation}\label{lambdapm4}
    \lambda_\pm=\frac{1}{4}\left(3\pm \sqrt{1-4m^2 \rho_0 ^2}\right)\,,
\end{equation}
and leads to the condition
\begin{equation}\label{condm2}
    m^2<\frac{1}{4\rho_0^2}\,.
\end{equation} 
On the other hand, using \eqref{IdGammasEuler} and \eqref{formR}, we can construct a vector $R_\mu$ entering in the Skyrme action \eqref{Skyrmeaction}. As shown in Section \ref{MYMtoS}, the corresponding scalar field $U$ will solve the Skyrme equations \eqref{skyrme}. When considering the metric \eqref{prodmanifold2}, Einstein equations \eqref{EeqMYM} can be put in the same reduced form given in \eqref{effEqs}, but in this case the constants $R_0$ and $\Lambda_{\rm eff}$ are given by
\begin{equation}\label{R0Lambda4}
\begin{aligned}
R_0&=2\Lambda-\frac{16G\lambda^{2}\left(\lambda-1\right)^{2}}{\gamma^{2}\rho_{0}^{4}}\,, \\
\Lambda_{\rm eff}&=\Lambda-\frac{\gamma^2-4Gm^2\lambda^2}{\gamma^2\rho_0^2}+\frac{8G\lambda^2\left(\lambda-1\right)^2}{\gamma^2\rho_0^4}.
\end{aligned}
\end{equation}
Solutions of these equations can be found exactly in the same way as done in the previous section.

\subsubsection*{D>4}
For $D>4$, the solutions of \eqref{effEqs} are product manifolds of $S^2$ and $(D-2)$-dimensional geometries $M_{D-2}$ with constant Ricci scalar $R_0$. The metric $h_{ab}$ satisfies Einstein equations in $D-2$ dimensions with an effective cosmological constant $\Lambda_{\rm eff}$ given in \eqref{R0Lambda4}.
\subsubsection*{D=4}
For $D=4$ the Einstein tensor $G_{ab}^{(h)}$ vanishes and the manifold $M_2$ given by a constant curvature space. Then \eqref{effEqs} leads to the condition $\Lambda_{\rm eff}=0$, whose solutions is given by
\begin{equation}\label{solrho4}
    \rho_{0\pm}^2=\frac{\gamma^2-4Gm^2\lambda^2\pm\sqrt{\left(\gamma^2-4Gm^2\lambda^2\right)^2-32\gamma^2G\lambda^2\left(\lambda-1\right)^2\Lambda}}{2\gamma^2\Lambda} \,.
\end{equation}
Using the anstaz \eqref{metrich} for the metric $h_{ab}$, the profile function $f(r)^2$ is found to have the form \eqref{fr} but now with $R_0$ given by Eq. \eqref{R0Lambda4}. Note that the solution \eqref{solrho4} can be obtained from the corresponding solution  found in $D=5$ by applying the following rescalings in Eq. \eqref{solrhopm}:
\begin{equation}\label{rescalings}
\begin{aligned}
& m_{(D=5)}^2\longrightarrow 2m_{(D=4)}^2 \,,\hskip1.5truecm
& G_{(D=5)}\longrightarrow  \mu G_{(D=4)} \,,\\
& \Lambda_{(D=5)} \longrightarrow 3 \Lambda_{(D=4)} \,,\hskip1.5truecm
& \gamma_{(D=5)}^2\longrightarrow  \mu\gamma_{(D=4)}^2 
 \,.
\end{aligned}
\end{equation}
where $\mu$ is a parameter with dimension of length required by dimensional analysis. Moreover, this rescaling of $m^2$ allows one to obtain the solution for $\lambda$ given in \eqref{lambdapm4} for $D=4$ from the corresponding solution \eqref{sollambda1} found in $D=5$. Similarly, the analysis for $\rho^2_{0\pm}$ given in Eqs. \eqref{5cond1} and \eqref{rhozero} also holds here when the rescalings \eqref{rescalings} are implemented.

\subsection{Euclidean solutions}\label{Euc}

Merons are usually understood as Euclidean solutions. So far, we have considered Meron-like Lorentzian solutions of the Einstein--MYM system. It is therefore natural to expect that these configurations admit an Euclidean continuation. This is indeed the case. As we will see, some of these solutions can have vanishing or finite Euclidean action, and therefore they represent saddle points in the path integral of both Einstein--MYM theory and in the Einstein--Skyrme model. The action \eqref{MYM} is brought to Euclidean signature by means of the standard Wick rotation $t=-i\tau$, $A_0=iA_4$, which leads to
\begin{equation}\label{EucEMYM}
I_{\rm Euclidean}=-\frac{1}{16\pi}\int d^D x ~\sqrt{g}\left\{\frac{1}{G}\left( R-2\Lambda\right)+\frac{1}{ \gamma^{2}}
{\rm Tr}\left[  F^{\mu\nu}F_{\mu\nu}+2m^{2}A^{\mu}A_{\mu}\right] \right\}\,.
\end{equation}
Now we proceed in complete analogy to \cite{Canfora:2017yio}. Since the Meron solutions satisfy the condition $A_4=0$, all the solutions that have been considered in the previous section are directly translated into solutions of the Euclidean theory by simply introducing the Euclidean time in the space-time metrics. Of particular interest is the example given in \eqref{finiteE} for $D=4$. In the Euclidean formulation correspond to a Meron gauge of the form given in \eqref{Aform} and \eqref{GammasEuler}, defined on the geometry
\begin{equation}\label{euc4d}
ds^{2}=d\tau^2 + \frac{\rho_0^{2}}{4}\left(d\psi^{2}+d\theta^{2}+d\phi^{2}+2\cos\theta d\phi d\psi \right) \,.
\end{equation} 
Solving Einstein equation leads to the same results found in Eqs. \eqref{lambdaRxS} and \eqref{rhoRxS} for $\Lambda$ and $\rho_0$, respectively. These relations, however, lead to a vanishing on-shell action
\begin{equation}
I_{\rm on-shell}=\frac{\pi\beta \rho_0}{4G}\left(\rho_0^2 \Lambda -3\right)+ \frac{6\pi\beta\lambda^2}{\gamma^2}\left(\frac{(\lambda -1)^2}{\rho_0}+\frac{m^2\rho_0}{4}\right) =0\,,
\end{equation}
which resembles the behaviour of gravitational instantons \cite{Eguchi:1978gw,Oliva:2009ip} and certain generalised Skyrme fieds \cite{Ferreira:2004np}.

On the other hand, as done in \cite{Canfora:2017yio}, in $D=3$ one can construct an Euclidean Meron on $S^3$. In this picture, one of the angles of the three-sphere, say $\psi$, is defined as a periodic Euclidean time with period $\beta=4\pi$:
\begin{equation}
ds^{2}=\frac{\rho_0^{2}}{4}\left(d\tau^{2}+d\theta^{2}+d\phi^{2}+2\cos\theta d\phi d\tau\right) \,.
\end{equation}
This leads to a $(2+1)$-dimensional massive Meron (or Skyrmion) solution with Euclidean action
\begin{equation}
I_{\rm on-shell}=\frac{\pi\rho_0}{4G}\left(\rho_0^2 \Lambda -3\right)+ \frac{6\pi\lambda^2}{\gamma^2}\left(\frac{(\lambda -1)^2}{\rho_0}+\frac{m^2\rho_0}{4}\right) \,.
\end{equation}
where Einstein equations fix $\rho_0$ to be
\begin{equation}\label{rhoEucS3}
    \rho^2_{0\pm}=\frac{\gamma^2-Gm^2\lambda^2 \pm\sqrt{\left(\gamma^2-Gm^2\lambda^2\right)^2+16\gamma^2G\Lambda \lambda^2\left(\lambda-1\right)^2}}{2\gamma^2\Lambda} \,.
\end{equation}
Unlike the previous case \eqref{euc4d}, here the cosmological constant is not constrained. This means that the Meron (Euclidean Skyrmion) on $S^3$ is a solution with finite on-shell action of the Euclidean Einstein-MYM theory (or the Euclidean Einstein--Skyrme model). One can see that for $m=0$ this result reduces to the one found in \cite{Canfora:2017yio} when restricted to the Euclidean Einstein--Yang--Mills case. Moreover, it is possible to generalise this result by adding a Chern-Simons term in the action. The Chern-Simons overall constant is imaginary in Euclidean signature, which leads to an imaginary value of $\lambda$.  As argued in \cite{Canfora:2017yio}, the solution that arises in this case define a complex saddle point, which have been shown to be necessary in order to have a well-defined path integral at the non-perturbative level \cite{Dunne:2016jsr,Dorigoni:2014hea}.

\section{Conclusions}
\label{conclusions}
We have studied Meron configurations in Einstein--MYM theory and we have shown that these solutions are in correspondence with hedgehog solitons in the Einstein--Skyrme model. This identifications of solutions is valid in any space-time dimensions. We have shown that implementing the Meron ansatz for the gauge connection $A=\lambda U^{-1}dU$ in the MYM equations leads to the the Skyrme equations for $U$. The identification requires to suitable identify the coupling constants of both theories, which also allows one to reduce the MYM energy-momentum to the corresponding Skyrme energy-momentum tensor. This last result further implies that Einstein equations in both system take the same form. Therefore, finding a gravitating Meron solution of Einstein--MYM theory automatically leads to the existence of a gravitating Skyrmion solution in the Einstein--Skyrme model. 

Next, we turned our attention to the $SU(2)$ case, where we have shown that, in order to have a solution of the MYM equations, the space-time geometries must be restricted to product manifolds of the form $M_{D-3} \times S^3$. Then, Einstein equations for D>5 can be reduced to an effective set of lower dimensional Einstein equations, while for $D=5$ the solution is given by the product of a constant curvature two-dimensional space-time and the three-sphere. Moreover, for $D=4$,  it is possible to construct a configuration with finite energy. In the Euclidean case, the analogue of this solution has vanishing on-shell action. Along these lines, the Euclidean formulation allows one to find a finite-action solution in D=3 by considering one of the direction of the three-sphere as a periodic imaginary time. An immediate consequence of this analysis is that our massive Merons/Skyrmions cannot be defined on Minkowski space and, therefore, they do not exist if gravity is turned off. Subsequently, We have shown how an identification on three sphere allows to define Merons on manifolds of the form $M_{D-2} \times S^2$. In this case the analysis of the solutions for $D>4$ and for $D=4$ follows in complete analogy to the previous case. In both type of manifolds, Merons are defined using the generalised hedgehog ansatz. In each case, the $SU(2)$ group element $U$ used to construct a gravitating Meron gauge field defines a Skyrmion when the Yang--Mills coupling constant and the gauge field mass $m^2$ are suitable identified with the coupling constants of the Skyrme model. In particular, we have shown that massive Merons can be defined on the manifolds (A)dS$_2\times S^3$ and (A)dS$_2\times S^2$. These manifolds are relevant in the study of near-horizon black hole geometries \cite{Maldacena:1998uz,Bardeen:1999px}. The solutions found in this article could thus be used to relate black holes a with Skyrmion hair \cite{Luckock:1986tr,Droz:1991cx,Heusler:1992av} to black holes a with MYM hair. We hope to address this question in a future work.

An interesting future direction is to extend these results to more general gauge groups. For example, exact solutions of the Skyrme model have been recently discovered in the $SU(3)$ case \cite{Ayon-Beato:2019tvu}. This scenario is physically sensible and relating solutions of the MYM equations with Skyrmions could have implications when studying non-perturbative aspects of QCD.

\subsection*{Acknowledgements}
We would like to thank R. Caroca, A. Gallego for helpful remarks and comments. We are specially indebted to F. Canfora for careful reading of the manuscript and for valuable discussions and suggestions throughout the development of this work. We acknowledge the referee for pointing out relevant comments, as well as for suggesting to consider the Euclidean continuation of our initial results. MI has been supported by FONDECYT grant N$^{\,\underline\circ}$ 3180594. PS-R acknowledges the School of Physics and Astronomy of the University of Leeds for hospitality and support as invited researcher.

\providecommand{\href}[2]{#2}\begingroup\raggedright\endgroup

\end{document}